\documentclass[12pt,letterpaper,twoside]{article}
\pdfoutput=1

\usepackage{natbib}
\usepackage{doi}
\usepackage{graphicx}
\usepackage{times}
\usepackage{amsmath}
\usepackage{amssymb}
\DeclareMathOperator{\sign}{sign}

\usepackage{url}

\usepackage{fancyhdr}
\fancypagestyle{firststyle}
{
   \fancyhf{}
   \fancyfoot[R]{arXiv:1905.02752v3 [cs.DM]} 
   \fancyfoot[L]{Copyright~\copyright~2019, Vincent A. Cicirello.}
}

\hypersetup{pdfinfo={
Title={Kendall Tau Sequence Distance: Extending Kendall Tau from Ranks to Sequences},
Author={Vincent A. Cicirello},
Subject={arXiv:1905.02752v3 [cs.DM], October 16, 2019},
Keywords={edit distance, Kendall tau, sequence distance, string distance}
}}

\begin{document}

\date{Report: October 16, 2019}

\title{\Large\bf Kendall Tau Sequence Distance: Extending Kendall Tau from Ranks to Sequences}

\author{Vincent A. Cicirello \\
  Computer Science \\
  Stockton University \\
  101 Vera King Farris Drive\\
  Galloway, NJ 08205\\
  \url{https://www.cicirello.org/}}

\maketitle

\thispagestyle{firststyle}

\begin{abstract}
An edit distance is a measure of the minimum cost sequence of edit operations
to transform one structure into another.  Edit distance is most commonly
encountered within the context of strings, where \citeauthor{wagner74}'s
string edit distance is perhaps the most well-known.  However, edit distance is not
limited to strings.  For example, there are several edit distance measures
for permutations, including \citeauthor{wagner74}'s string edit distance since a
permutation is a special case of a string.  However, another
edit distance for permutations
is Kendall tau distance, which is the number of pairwise element
inversions.  On permutations, Kendall tau distance is 
equivalent to an edit distance with adjacent swap as the edit operation.
A permutation is often used to represent a total ranking over a set of elements.
There exist multiple extensions of Kendall tau distance from total rankings (permutations)
to partial rankings (i.e., where multiple elements may have the same rank), but none of
these are suitable for computing distance between sequences.
We set out to explore extending Kendall tau distance in a different direction, namely
from the special case of permutations to the more general case of strings or sequences
of elements from some finite alphabet.  We name our distance metric 
Kendall tau sequence distance, and define it as the minimum number of adjacent swaps
necessary to transform one sequence into the other.  We provide two $O(n \lg n)$ 
algorithms for computing it, and experimentally compare their relative performance.  
We also provide reference implementations of both algorithms in an open source Java library.
\end{abstract}

\section{Introduction}

There exists a wide variety of metrics for computing the distance between 
permutations~\citep{ronald1995,ronald1997,ronald1998,fagin2003,campos2005,marti2005,sevaux2005,sorensen07,meila2010,cicirello2013,cicirello2016,cicirello2018,cicirello2019}.
The different permutation metrics that are available consider different characteristics
of the permutation depending upon what it represents (e.g., a mapping between
two sets, a ranking over the elements of a set, or a path through a graph). 
There is at least one instance where a metric on strings is suggested for permutations.
Specifically, \citet{sorensen07} suggested using string edit distance to measure distance
between permutations.
The specific edit distance suggested by \citeauthor{sorensen07} 
was the string edit distance of \citet{wagner74}.
In general, the edit distance between two structures is the minimum cost sequence of
edit operations to transform one structure into the other.
\citeauthor{wagner74}'s string edit distance
is the minimum cost sequence of edit operations
to transform one string into the other where the edit operations are element removals,
insertions, and replacements.  
The usual algorithm for computing it is the dynamic programming
algorithm of \citet{wagner74}, which has a runtime of $O(n*m)$ where $n$ and $m$ are the lengths
of the strings (in the case of permutations, runtime is $O(n^2)$ since lengths are the same).

In this paper, we begin with a metric on permutations, and adapt it
to measure the distance between sequences (i.e., strings, arrays, or any other sequential data).
The specific metric that we adapt to sequences is Kendall tau distance.  Kendall tau distance is
a metric defined for permutations that is itself an adaptation of Kendall tau rank correlation~\citep{kendall1938}.
As a metric on permutations, Kendall tau distance assumes that a permutation represents a ranking over some set
(e.g., an individual's preferences over a set of songs or books, etc), and is the count of the number of pairwise
element inversions.
We review Kendall tau distance, for permutations, in Section~\ref{sec:tau}, along with existing extensions
for handling partial rankings (i.e., instead of a permutation or total ordering, 
partial orderings with tied ranks are compared).  

In the case of permutations, where each element of the set is represented
exactly one time in each permutation, Kendall tau distance is the minimum number of adjacent swaps
necessary to transform one permutation into the other.  Thus, in the case of permutations, Kendall tau
distance is an edit distance where the edit operations are adjacent swaps.  
Due to this relationship,
it is sometimes referred to as bubble sort distance, 
since bubble sort functions via adjacent element swaps.
However, as soon as you leave the realm of permutations, 
existing forms of Kendall tau no longer correspond to
an adjacent swap edit distance.  We provide an example of this in Section~\ref{sec:example}.

In the case of comparing partial rankings, the existing extensions 
of Kendall tau distance to partial rankings are fine.
However, if we are comparing sequences (e.g., strings, arrays of data points, etc)
that do not represent a ranking, then the partial ranking versions of 
Kendall tau distance do not apply.
We propose a new extension of Kendall tau distance for sequences in Section~\ref{sec:tauseq}.
We call it Kendall tau sequence distance, and show that it meets the requirements of a metric.
It is applicable for computing the distance between pairs of sequences, where both 
sequences are of the same length, and consist in the same set of elements (i.e., 
duplicates are allowed, but both sequences must have the same duplicated elements).
It is otherwise applicable to strings over any alphabet or any other form of 
sequence (such as an array of integers or an array of floating-point values, etc).  
We argue that it is more relevant as a measure of array sortedness than the 
existing partial ranking adaptations of Kendall tau. 
In Section~\ref{sec:algs}, we provide two $O(n \lg n)$ algorithms for computing Kendall tau sequence
distance.

We implemented both algorithms in Java, and we have added those reference implementations 
to JavaPermutationTools (JPT), an open source Java library of data structures 
and algorithms for computation on permutations
and sequences~\cite{cicirello2018}, which can be found at \url{https://jpt.cicirello.org/}.
In Section~\ref{sec:experiments}, we experimentally compare the relative performance of the
two algorithms.  The code to replicate these experiments is also available in the
code repository of the JPT.

\section{Kendall tau distance for permutations}\label{sec:tau}

\subsection{Notation}\label{sec:notation}

Without loss of generality, we will assume a permutation of length $n$ is a permutation of the integers in the set $S={1, 2, \ldots, n}$.
Let $\sigma(i)$, where $i \in S$, be the position of element $i$ in the permutation $\sigma$.  If the permutation is a ranking over a
set of $n$ objects, then $\sigma(i)$ represents the rank of object $i$ in that ranking.  Let $p(r)$, where $r \in S$, be the element
in position $r$ of the permutation (or with rank $r$).  Our notation assumes that the index into the permutation begins at 1.  

The $\sigma$ and $p$ are two alternative
representations of the permutation.  They are related as follows:
$\sigma(i) = r \iff p(r) = i$.  Throughout the paper, we will use whichever is more convenient in the given context.

We will initially assume that permutations (whether defined with $\sigma$ or with $p$) are true
permutations.  That is, we assume $\sigma(i)=\sigma(j) \iff i=j$ and also that $p(r_1)=p(r_2) \iff r_1=r_2$.  
Therefore, if the application is
one of rankings, we assume that there are no ties.  
In other words, two objects
have the same rank only if they are the same object; and each object has only one rank.
We relax this assumption later in Section~\ref{sec:partial}.

\subsection{Kendall tau rank correlation}\label{sec:tau-rank}

Kendall tau distance for permutations is strongly based on the Kendall tau rank correlation coefficient.
Consider two permutations $\sigma_1$ and $\sigma_2$.
The Kendall tau rank correlation coefficient~\citep{kendall1938} is defined as:
\begin{equation}
\tau(\sigma_1, \sigma_2) = \frac{2}{n*(n-1)} \sum_{i,j \in S \land i < j} \sign(\sigma_1(i) - \sigma_1(j)) * \sign(\sigma_2(i) - \sigma_2(j)) .
\end{equation}
The summation has a maximum value of $n*(n-1)/2$, which occurs when $\sigma_1 = \sigma_2$; and 
the summation has a minimum value of $-n*(n-1)/2$, which occurs when $\sigma_1$ is the reverse of $\sigma_2$.
The $2/(n*(n-1))$ term scales such that $\tau \in [-1, 1]$.

Another way of expressing it is as follows:
\begin{equation}
\tau(\sigma_1, \sigma_2) = \frac{2}{n*(n-1)} ( |C| - |D| ),
\end{equation}
where $C$ is the set of concordant pairs, defined as:
\begin{equation}
C = \{(i,j) \in S \times S \, | \, i < j \land ( \sigma_1(i) < \sigma_1(j) \land \sigma_2(i) < \sigma_2(j) \lor \sigma_1(i) > \sigma_1(j) \land \sigma_2(i) > \sigma_2(j))\} ,
\end{equation}
and $D$ is the set of discordant pairs:
\begin{equation}\label{eq:D}
D = \{(i,j) \in S \times S \, | \, i < j \land ( \sigma_1(i) < \sigma_1(j) \land \sigma_2(i) > \sigma_2(j) \lor \sigma_1(i) > \sigma_1(j) \land \sigma_2(i) < \sigma_2(j))\} .
\end{equation}

\subsection{Kendall tau distance}\label{sec:tau-dist}

For a function $d : S \times S \rightarrow \mathbb{R}$ to be a measure of distance, we must have non-negativity ($d(i,j) \geq 0$ for all $i,j\in S$),
identity of indiscernibles ($d(i,j)=0 \iff i=j$ for all $i,j\in S$), and symmetry ($d(i,j)=d(j,i)$ for all $i,j\in S$).
Further for $d : S \times S \rightarrow \mathbb{R}$ to be a metric, 
it must also satisfy the triangle inequality ($d(i,j) \leq d(i,k) + d(k,j)$ for all $i,j,k\in S$).
The Kendall tau rank correlation coefficient is not a measure of distance (e.g., it clearly doesn't satisfy the first two
requirements of non-negativity and identity of indiscernibles.

Kendall tau distance (for permutations) is found in the literature in two forms, as follows:
\begin{equation}\label{eq:K}
K(\sigma_1, \sigma_2) = |D|,
\end{equation}
and
\begin{equation}
K(\sigma_1, \sigma_2) = \frac{2 |D|}{n*(n-1)},
\end{equation}
where $D$ is the set of discordant pairs as previously defined in Equation~\ref{eq:D}.
The only difference between these is that in the latter case, the distance is normalized to lie in the interval $[0,1]$,
and in the former case the distance lies in the interval $[0,n*(n-1)/2]$.
We have $K(\sigma_1, \sigma_2) = 0$ only when $\sigma_1 = \sigma_2$.  And the maximum occurs when 
$\sigma_1$ is the reverse of $\sigma_2$. 
Kendall tau distance for permutations satisfies all of the metric properties.  

The version seen in Equation~\ref{eq:K}
is also equal to the minimum
number of adjacent swaps necessary to transform one permutation $p_1$ into the other permutation $p_2$.  
That is, it is an edit distance where the edit operation is adjacent swap.  
Consider as an example, the permutations $\sigma_1 = [ 2, 4, 1, 3 ]$
and $\sigma_2 = [ 4, 1, 3, 2 ]$.    
Their equivalents in the other notation are $p_1 = [ 3, 1, 4, 2 ]$ and
$p_2 = [ 2, 4, 3, 1 ]$.  The discordant pairs are $D = \{ (1, 2), (1, 4), (2, 3), (2, 4), (3, 4) \}$.
Thus, $K(\sigma_1, \sigma_2) = 5$ in this example.  
You can transform $p_1 = [ 3, 1, 4, 2 ]$ into $p_2$ via the following sequence of five adjacent swaps:
$[ 3, 4, 1, 2 ]$, $[ 3, 4, 2, 1 ]$, $[ 4, 3, 2, 1 ]$, $[ 4, 2, 3, 1 ]$, $[ 2, 4, 3, 1 ] = p_2$.
You cannot do it with fewer than five adjacent swaps in this example. 

Note that as an adjacent swap edit distance, it specifically concerns the $p$ representation of the permutation and
not the $\sigma$ notation.  For example, adjacent swaps on $\sigma_1$ leads to a shorter sequence (3 swaps):
$[ 4, 2, 1, 3 ]$, $[ 4, 1, 2, 3 ]$, $[ 4, 1, 3, 2 ] = \sigma_2$. 
However, there is an equivalent operation for the $\sigma$ notation, swapping consecutive ranks (i.e., rank 1 with 2,
2 with 3, etc).  That is, since $p$ lists the elements in their ``ranked'' order, an adjacent swap in $p$ is
equivalent to exchanging the ranks of two elements whose ranks differ by 1.  

Another (slightly less direct) 
way of connecting the $\sigma$ representations of the permutations to the view of Kendall tau distance as
an adjacent swap edit distance leads to the common $O(n \lg n)$ algorithm for computing it.
Define the following list of ordered pairs: 
\begin{equation}
T = [ (\sigma_1(1), \sigma_2(1)), (\sigma_1(2), \sigma_2(2)), \ldots, (\sigma_1(n), \sigma_2(n)) ] .
\end{equation}
Sort $T$ by first component of the tuples 
(any sorting algorithm will do, but preferably one with worst case runtime in $O(n \lg n)$).
Let $T'$ be the sorted $T$.  While sorting $T'$ by the second component (e.g., such as by mergesort),
count the number of inversions.  The number of inversions in $T'$ (per second components of tuples) is
the Kendall tau distance, and is the number of adjacent swaps necessary to sort $T'$.
For the previous example where we had $\sigma_1 = [ 2, 4, 1, 3 ]$
and $\sigma_2 = [ 4, 1, 3, 2 ]$, we define $T = [ (2,4), (4,1), (1,3), (3,2) ]$.
Sorting by first component results in $T' = [ (1,3), (2,4), (3,2), (4,1) ]$, which has
5 inversions (per second components of tuples): 3 with 2, 3 with 1, 4 with 2, 4 with 1, and 2 with 1.
This $O(n \lg n)$ approach to computing Kendall tau distance has been described by several
previously, such as by \citet{knight66} though in the context of Kendall tau rank correlation.

\subsection{Partial ranking Kendall tau distance}\label{sec:partial}

We now amend the notation previously introduced in Section~\ref{sec:notation}.  Specifically, we will now 
assume that rankings may be partial (i.e., there may be ties).  That is, although $i=j \implies \sigma(i)=\sigma(j)$
is still the case, we now allow $\sigma(i)=\sigma(j)$ in cases where $i \neq j$ (i.e., two different elements may have same rank).

The simplest way to extend Kendall tau rank correlation or Kendall tau distance
to partial rankings is to compute it without modification.  That is, compute the number of discordant pairs, etc
and use the definitions of Sections~\ref{sec:tau-rank} and~\ref{sec:tau-dist}.
The algorithm of \citet{knight66}
described in the previous section is actually specified to handle partial rankings in this way.
In the first sort, where the list of tuples $T$ is sorted by the first component of the tuples,
\citet{knight66} indicates to break ties using the second component.

Among the potential problems with directly applying Kendall tau distance
without modification to partial rankings is that it no longer
meets the metric properties.  \citet{fagin06} developed the $K^{(p)}$,
known as the Kendall distance with penalty parameter $p$ to deal with this, and determined the
range of values for the penalty parameter that enables fulfilling the metric properties.
Define $K^{(p)}$ as follows:
\begin{equation}\label{eq:Kp}
K^{(p)}(\sigma_1,\sigma_2) = |D| + p * |E| ,
\end{equation}
where $D$ is still the set of discordant pairs, as previously defined in Equation~\ref{eq:D}.  Note the strict $<$ and $>$
in the definition of $D$, and that a tie within either permutation is not a discordant pair.
$E$ is the set of pairs that are ties in one permutation, but not the other (i.e., one ranking
considers the objects equivalent, but the other does not).  Therefore, $E$ is defined as:
\begin{equation}
E = \{(i,j) \in S \times S \, | \, i < j \land ( \sigma_1(i) = \sigma_1(j) \land \sigma_2(i) \neq \sigma_2(j) \lor \sigma_1(i) \neq \sigma_1(j) \land \sigma_2(i) = \sigma_2(j))\} .
\end{equation}
\citet{fagin06} showed that $K^{(p)}$ is a metric when $0.5 \leq p \leq 1$, and that it is what they termed a ``near metric'' when $0 < p < 0.5$,
and that it is not a distance when $p=0$.  We do not use their ``near metric'' concept here so we leave it 
to the interested reader to consult \citet{fagin06}.

\subsection[Partial ranking Kendall tau distance is not adjacent swap edit distance]{Partial ranking Kendall tau distance $\neq$ adjacent swap edit distance}\label{sec:example}

As a distance metric on partial rankings, the Kendall distance with penalty parameter $p$ of \citet{fagin06}
is an effective choice, and commonly used in the context of comparing partial rankings.
However, it is not adjacent swap edit distance.  Consider the following illustrative example.
Let $\sigma_1 = [ 1, 2, 3, 1, 1, 2, 2 ]$ and $\sigma_2 = [ 3, 2, 1, 2, 1, 2, 1 ]$.  
In this case, the set of discordant pairs is $D = \{ (1, 2), (1, 3), (1, 6), (1, 7), (2, 3), (3, 4), (3, 6), (4, 7) \}$,
and the set $E= \{ (1, 4), (1, 5), (2, 4), (2, 7), (3, 5), (3, 7), (4, 5), (4, 6), (5, 7), (6, 7) \}$.
Thus, $K^{(p)}(\sigma_1,\sigma_2) = 8 + 10 p$ (Equation~\ref{eq:Kp}).

You can compute $|D|$ and $|E|$ without actually computing the sets $D$ and $E$ via the approach of \citet{knight66}
based on sorting.  Let $T = [ (1, 3), (2, 2), (3, 1), (1, 2), (1, 1), (2, 2), (2, 1) ]$.
Sort $T$ by first component of tuples, breaking ties via the second components, and obtain:
$T' = [ (1, 1), (1, 2), (1, 3), (2, 1), (2, 2), (2, 2), (3, 1) ]$.
You can finally sort $T'$ via mergesort (or another $O(n \lg n)$ sort), with the sort modified to count
inversions.  In this case, there are 8 inversions in $T'$, which is equal to $|D|$.  It is also straightforward
enough to compute $|E|$.  

The $|D|$ in this example is the minimum number of adjacent swaps necessary to sort $T'$.  However, it is not
the minimum number of adjacent swaps necessary to transform $\sigma_1$ into $\sigma_2$.
That can be done with fewer than eight adjacent swaps.  Specifically, it can be done via the following sequence
of six adjacent swaps: $\sigma_1 = [ 1, 2, 3, 1, 1, 2, 2 ]$, $[ 1, 3, 2, 1, 1, 2, 2 ]$, $[ 3, 1, 2, 1, 1, 2, 2 ]$, 
$[ 3, 2, 1, 1, 1, 2, 2 ]$, $[ 3, 2, 1, 1, 2, 1, 2 ]$, $[ 3, 2, 1, 2, 1, 1, 2 ]$, $[ 3, 2, 1, 2, 1, 2, 1 ] = \sigma_2$.

Now, previously in Section~\ref{sec:tau-dist}, we saw that with full rankings (i.e., permutations)
Kendall tau distance is equal to the minimum number of adjacent swaps to transform $p_1$ into $p_2$
(i.e., an adjacent swap edit distance on the $p$ notation, where $p(r)$ yields the object
with rank $r$).  With partial rankings, we don't have the equivalent of $p$ since multiple objects
may have the same rank.  One attempt might be to allow $p(r)$ to map to the set of objects with
rank $r$.  Thus, for the example of the prior paragraph, we'd have $p_1 = [\{1, 4, 5\}, \{2, 6, 7\}, \{3\}]$,
and $p_2 = [\{3, 5, 7\}, \{2, 4, 6\}, \{1\}]$.  Transforming $p_1$ to $p_2$ via adjacent swaps (if we define an adjacent swap
in this context as swapping two elements in adjacent sets) can be done with four such swaps.  

Also in Section~\ref{sec:tau-dist}, for full rankings, we saw that Kendall tau distance is equal to the minimum number
of applications of an operation that exchanges the ranks of two elements whose ranks differ by 1.
For this example, a sequence of four such operations can transform $\sigma_1 = [ 1, 2, 3, 1, 1, 2, 2 ]$
into $\sigma_2 = [ 3, 2, 1, 2, 1, 2, 1 ]$.  That sequence is as follows: 
$\sigma_1 = [ 1, 2, 3, 1, 1, 2, 2 ]$, $[ 1, 2, 3, 2, 1, 2, 1 ]$, $[ 1, 3, 2, 2, 1, 2, 1 ]$, $[ 2, 3, 1, 2, 1, 2, 1 ]$, $[ 3, 2, 1, 2, 1, 2, 1 ] = \sigma_2$.
This is equivalent to our redefinition of $p(r)$ to the set of elements with rank $r$.

There is no interpretation where $K^{(p)}$ or any other partial ranking variation of Kendall tau distance
that is based on the number of discordant pairs is equivalent to an adjacent swap edit distance.
The example of this section illustrates this in that there are eight discordant pairs (thus $K^{(p)} \geq 8$ unless
$p$ is negative) while less than eight adjacent swaps is sufficient for sorting the permutation
(either 6 or 4 depending upon the interpretation of ``adjacent swap'' and the representation to which it is applied).

\subsection{Positions of elements in a sequence are not ranks}

If the sequences we are comparing do not define rankings, then the partial ranking variants of Kendall tau distance
are not applicable as it would be arbitrary to impose a ranking interpretation upon them, and also likely
to lead to a nonsensical interpretation.  For example, consider the string $s$: ``abacab''.  It would be
arbitrary to impose a lexicographical order of the characters as if they are ranks (e.g., ``a'' as 1, ``b'' as 2, etc),
such as transforming $s$ to $\sigma = [1, 2, 1, 3, 1, 2]$.  Or, if you consider position in the sequence to be an element's rank,
then you'd have something meaningless like ``a'' is simultaneously ranked first, third, and fifth.

\section{Kendall tau sequence distance}\label{sec:tauseq}

\subsection{Notation}\label{sec:seqNotation}

Let $s$ be a sequence of length $n$, where $s(i) \in \Sigma$ for some alphabet $\Sigma$ and $i \in \{ 0, 1, \ldots, n-1\}$.
The alphabet $\Sigma$ can be a character set for some language, but can also be the set of integers, the set of real numbers,
the set of complex numbers, or any other set of elements.  The alphabet $\Sigma$ is not necessarily a finite alphabet,
although we do assume finite length sequences (i.e., $n$ is finite).

Without loss of generality, we also assume that the elements of the alphabet $\Sigma$ can be ordered.  The specific ordering
does not affect the measure of distance between the sequences.

\subsection[Kendall tau sequence distance equals adjacent swap edit distance]{Kendall tau sequence distance $=$ adjacent swap edit distance}

We previously saw in Section~\ref{sec:tau-dist} that the original form of Kendall tau permutation distance
is equivalent to an adjacent swap edit distance when applied to permutations (i.e., no duplicates) and specifically
when applied to the $p$ representation (and not the $\sigma$ representation).
But that the existing extensions of Kendall tau beyond permutations (e.g., partial ranking variants)
are not equivalent to an adjacent swap edit distance.

We now define the Kendall tau sequence distance, $\tau_S$, as follows:
\begin{equation}\label{eq:tauS}
\tau_S(s_1,s_2) = \text{minimum number of adjacent swaps that transforms $s_1$ into $s_2$}.
\end{equation}
where $s_1$ and $s_2$ are sequences as defined in Section~\ref{sec:seqNotation}.  We require
the lengths of the sequences to be equal, i.e., $|s_1| = |s_2|$.  And for each character $c \in \Sigma$,
we require $\text{count}(s_1, c) = \text{count}(s_2, c)$, where $\text{count}(s, c)$ is the number
of times that $c$ appears in $s$.  The $\tau_S$ distance is undefined if these conditions do not hold
for a specific pair of sequences.

The $\tau_S$ distance satisfies all of the metric properties.  It clearly satisfies non-negativity,
identity of indiscernibles, and symmetry.  We must have $\tau_S(s_1,s_2) \geq 0$, since it is not 
possible to apply a negative number of swaps.  If $s_1 = s_2$, then $\tau_S(s_1,s_2) = 0$ since 0
swaps are required to transform a sequence to itself.  And if $\tau_S(s_1,s_2) = 0$, then $s_1 = s_2$
since the only case when a sequence can be transformed to another with 0 adjacent swaps is obviously 
when the two sequences are identical.  It is also obvious that $\tau_S(s_1,s_2)=\tau_S(s_2,s_1)$.

The $\tau_S$ also satisfies the remaining metric property, the triangle inequality:
\begin{equation}
\tau_S(s_1,s_2) \leq \tau_S(s_1,s_3) + \tau_S(s_3,s_2) .  
\end{equation}
The proof is as follows (via contradiction).
Suppose there exists sequences $s_1$, $s_2$, and $s_3$, such that: $\tau_S(s_1,s_2) > \tau_S(s_1,s_3) + \tau_S(s_3,s_2)$.
The minimum cost edit sequence from $s_1$ to $s_3$ is $\tau_S(s_1,s_3)$ (by definition via Equation~\ref{eq:tauS}).
Likewise, the minimum cost edit sequence from $s_3$ to $s_2$ is $\tau_S(s_3,s_2)$.  One sequence of
edit operations that will transform $s_1$ to $s_2$ is to first transform $s_1$ to $s_3$, and then to transform
$s_3$ to $s_2$.  The cost of that edit sequence is clearly the sum of the costs of the two portions: 
$\tau_S(s_1,s_3) + \tau_S(s_3,s_2)$.  The minimum cost edit sequence to transform $s_1$ to $s_2$
must therefore be no greater than $\tau_S(s_1,s_3) + \tau_S(s_3,s_2)$, a contradiction.

\subsection[Two algorithms to compute Kendall tau sequence distance]{Two $O(n \lg n)$ algorithms to compute $\tau_S$}\label{sec:algs}

In this section, we present two $O(n \lg n)$ algorithms for computing $\tau_S$.
Both rely on an observation related to the optimal sequence of adjacent swaps for editing
one sequence $s_1$ to the other $s_2$, and specifically concerning duplicate elements.
If a mapping between the elements of $s_1$ and $s_2$ is defined, such that an element is mapped
to its corresponding position if the optimal sequence of adjacent swaps is performed, then
an element that appears only once in $s_1$ will be mapped to the only occurrence in $s_2$.
Furthermore, in such a mapping, if an element appears multiple times, then the $k$-th 
occurrence in $s_1$ will be mapped to the $k$-th occurrence in $s_2$.  For example,
consider $s_1 = [a, b, a, c, a, d, a]$ and $s_2 = [b, c, a, a, a, a, d]$. 
The elements that appear only once obviously map to their corresponding element in the 
other sequence, in this case: $s_1[1]$ to $s_2[0]$, $s_1[3]$ to $s_2[1]$, and $s_1[5]$ to $s_2[6]$.
In this example, however, there are also four copies of the element $a$.  The optimal
edit sequence of adjacent swaps must map them as follows: $s_1[0]$ to $s_2[2]$, $s_1[2]$ to $s_2[3]$,
$s_1[4]$ to $s_2[4]$, and $s_1[6]$ to $s_2[5]$.  Any other mapping would result in extra adjacent swaps
that cause two copies of element $a$ to pass each other.  For example, consider this 
sequence, $s = [b, c, a, a, d, e]$.  Swapping the two copies of element $a$ results in the same sequence.
In general, a swap of adjacent identical copies of the same element does not change the sequence, but accrues
a cost of 1.  

The two algorithms both generate a mapping of the indices of one sequence that correspond to the elements
of the other, as described above.  The two algorithms differ in how they generate the mapping.
The mapping, once generated, is a permutation of the integers in $\{0, 1, \ldots, n-1\}$.  And the $\tau_S$
is the number of permutation inversions in that mapping.

\subsubsection{Algorithm 1}

The first of two algorithms for computing $\tau_S$ is found in Figure~\ref{fig:alg1}.
Line 4 generates a sorted copy, $S$, of one of the two sequences.  This step can be implemented
with mergesort or another $O(n \lg n)$ sorting algorithm for a cost of $O(f_c(m)\, n \lg n)$,
where $f_c(m)$ is the cost of comparing sequence elements of size $m$.  If the sequences contain
primitive values, such as ASCII or Unicode characters, then $f_c(m) = O(1)$.  I have included the $f_c(m)$
term to cover the more general case of sequences of objects of any type.
Lines 5--11 uses $S$ to generate a
mapping $M$ between unique sequence elements to the integers in $\{0,1,\ldots,k-1\}$, where there
are $k$ unique characters appearing in the sequences.  The cost to generate this mapping is $O(f_c(m)\,n)$.

\begin{figure}[tp]
\centering
\begin{minipage}{4in}
\begin{tabbing}
$\tau_S(s_1, s_2)$ \= \\
1. \> \textbf{if} \= $|s_1| \neq |s_2|$ \\
2. \> \> \textbf{return} error: unequal length sequences \\
3. \> Let $n = |s_1|$ \\
4. \> Let $S$ be a sorted copy of $s_1$ \\
5. \> Let $M$ be a new array of length $n$ \\
6. \> $M[0] \leftarrow 0$ \\
7. \> \textbf{for} \= $i = 1$ \textbf{to} $n-1$ \textbf{do} \\
8. \> \> \textbf{if} \= $S[i] = S[i-1]$ \\
9. \> \> \> $M[i] \leftarrow M[i-1]$ \\
10. \> \> \textbf{else} \\
11. \> \> \> $M[i] \leftarrow M[i-1] + 1$ \\  
12. \> Let $B_1$ and $B_2$ be arrays of length $M[n-1] + 1$ of initially empty queues \\
13. \> \textbf{for} \= $i = 0$ \textbf{to} $n-1$ \textbf{do} \\
14. \> \> Let $j$ be an index into $S$, such that $S[j] = s_1[i]$.\\
15. \> \> Let $k$ be an index into $S$, such that $S[k] = s_2[i]$.\\
16. \> \> \textbf{if} \= $k$ is undefined \\
17. \> \> \> \textbf{return} error: sequences contain different elements \\
18. \> \> Add $i$ to the tail of queue $B_1[M[j]]$.\\
19. \> \> Add $i$ to the tail of queue $B_2[M[k]]$.\\
20. \> Let $P$ be an array of length $n$ \\
21. \> \textbf{for} \= $i = 0$ \textbf{to} $M[n-1]$ \textbf{do} \\
22. \> \> \textbf{if} \= lengths of queues $B_1[i]$ and $B_2[i]$ are different \\ 
23. \> \> \> \textbf{return} error: sequences contain different number of copies of an element \\
24. \> \> \textbf{while} \= queue $B_1[i]$ is not empty \textbf{do} \\
25. \> \> \> Remove the head of queue $B_1[i]$ storing it in $h_1$. \\
26. \> \> \> Remove the head of queue $B_2[i]$ storing it in $h_2$. \\
27. \> \> \> $P[h_1] \leftarrow h_2$ \\
28. \> Let $I$ be the number of inversions in $P$.\\
29. \> \textbf{return} $I$
\end{tabbing}
\end{minipage}
\caption{Algorithm for computing $\tau_S$}\label{fig:alg1}
\end{figure}

Lines 12--19 performs bucket sorts of $s_1$ and $s_2$ as follows.
It places index $i$ of $s_1$ into the bucket corresponding to the integer from
the mapping $M$ that represents character $s_1[i]$.  This requires a search of $S$
in step 14, which can be implemented with binary search in $O(f_c(m)\, \lg n)$ time since
$S$ is in sorted order.  The buckets are represented with queues to easily maintain
the order that duplicate copies of an element appear in the original sequence.
Adding to the tail of a queue is a constant time operation.  $B_1$ is an array of the buckets for
$s_1$.  In a similar manner, a bucket sort of $s_2$ is performed, and $B_2$ is an array of the
corresponding buckets.  The block in lines 12--19 has a total cost of $O(f_c(m)\, n \lg n)$
since the loop of line 13 iterates $n$ times and the binary searches in lines 14 and 15 
have a runtime of $O(f_c(m)\, \lg n)$.

Lines 20--27 iterates over the buckets, mapping the elements of $s_2$ to the corresponding elements
of $s_1$.  The resulting mapping is a permutation $P$ of the integers in $\{0, 1, \ldots, n-1\}$.
Where there are duplicates of a specific character of the alphabet $\Sigma$, they are mapped 
in the order of appearance.  For example, if character $c$ appears in positions 2, 5, 18 of $s_1$
and in positions 4, 7, 22 of $s_2$, then the permutation $P$ will have the following corresponding
entries: $P[2]=4, P[5]=7, P[18]=22$.  The nested loops in lines 21 and 24 iterate exactly one time
for each sequence index, i.e., a total of $n$ executions of the body (lines 25--27) of the nested loops.
The body of which contains only constant time operations.  Thus, the runtime of lines 20--27
is $O(n)$.

Counting permutation inversions (line 28) is done in $O(n \lg n)$ time with a modified
mergesort. 

The runtime of this first algorithm is therefore $O(f_c(m)\, n \lg n)$ due to the sort in line 4,
and the block of lines 12--19.  This is worst case as well as average case.
If the sequences contain values of a primitive type, such as ASCII or Unicode 
characters, primitive integers,
primitive floating-point numbers, etc, then $f_c(m) = O(1)$, and thus the runtime
of the algorithm simplifies to $O(n \lg n)$.

\subsubsection{Algorithm 2}

Our second algorithm for computing $\tau_S$ is found in Figure~\ref{fig:alg2}.
It is similar in function to the first algorithm, but generates the mapping
from unique sequence elements to integers differently.  Specifically, it uses a 
hash table, $H$ (initialized in line 4).  Lines 5--9 populates that hash table.
The loop in that block iterates $n$ times, and assuming the sequences contain 
elements of a primitive type then all operations in its body
can be implemented in constant time (e.g., the key check in line 7, and
the put in line 8 can be implemented in $O(1)$ time with
a sufficiently large hash table size).  Our implementation ensures that the load factor
of the hash table never exceeds 0.75 in order to achieve the constant number
of hashes.  Thus, the runtime of this block is $O(n)$ for sequences of primitive elements.
Otherwise, in general, it is $O(f_h(m)\, n)$ where $f_h(m)$ is the cost to hash an object 
of size $m$.

\begin{figure}[tp]
\centering
\begin{minipage}{4in}
\begin{tabbing}
$\tau_S(s_1, s_2)$ \= \\
1. \> \textbf{if} \= $|s_1| \neq |s_2|$ \\
2. \> \> \textbf{return} error: unequal length sequences \\
3. \> Let $n = |s_1|$ \\
4. \> Let $H$ be an initially empty hash table mapping sequence elements to integers. \\
5. \> $q \leftarrow 0$ \\
6. \> \textbf{for} \= $i = 0$ \textbf{to} $n-1$ \textbf{do} \\
7. \> \> \textbf{if} \= $s_1[i] \notin \text{keys}(H)$ \\
8. \> \> \> Put the mapping $(s_1[i], q)$ in $H$. \\
9. \> \> \> $q \leftarrow q + 1$ \\
10. \> Let $B_1$ and $B_2$ be arrays of length $q$ of initially empty queues \\
11. \> \textbf{for} \= $i = 0$ \textbf{to} $n-1$ \textbf{do} \\
12. \> \> $j \leftarrow H[s_1[i]]$ \\
13. \> \> $k \leftarrow H[s_2[i]]$ \\
14. \> \> \textbf{if} \= $k$ is undefined \\
15. \> \> \> \textbf{return} error: sequences contain different elements \\
16. \> \> Add $i$ to the tail of queue $B_1[j]$.\\
17. \> \> Add $i$ to the tail of queue $B_2[k]$.\\
18. \> Let $P$ be an array of length $n$ \\
19. \> \textbf{for} \= $i = 0$ \textbf{to} $q-1$ \textbf{do} \\
20. \> \> \textbf{if} \= lengths of queues $B_1[i]$ and $B_2[i]$ are different \\ 
21. \> \> \> \textbf{return} error: sequences contain different number of copies of an element \\
22. \> \> \textbf{while} \= queue $B_1[i]$ is not empty \textbf{do} \\
23. \> \> \> Remove the head of queue $B_1[i]$ storing it in $h_1$. \\
24. \> \> \> Remove the head of queue $B_2[i]$ storing it in $h_2$. \\
25. \> \> \> $P[h_1] \leftarrow h_2$ \\
26. \> Let $I$ be the number of inversions in $P$.\\
27. \> \textbf{return} $I$
\end{tabbing}
\end{minipage}
\caption{A second algorithm for computing $\tau_S$}\label{fig:alg2}
\end{figure}

Lines 10--17 is the bucket sort described in the previous algorithm.  However,
unlike Algorithm 1 which requires binary searches of a sorted array, Algorithm 2
instead relies on hash table lookups (lines 12--13) which can be implemented in
$O(1)$ time for primitive elements, or $O(f_h(m))$ time more generally.  
Thus, this block's runtime is $O(f_h(m)\, n)$, or $O(n)$ for sequences of primitive elements.

Lines 18--25 iterates over the buckets, as in Algorithm 1, to generate the
permutation mapping elements between the two sequences.  It is unchanged from
Algorithm 1, and thus has a runtime of $O(n)$.

Line 26 counts permutation inversions, just like in Algorithm 1, and thus
has a runtime of $O(n \lg n)$.

The runtime of Algorithm 2 is thus $O(f_h(m)\, n + n \lg n)$.  For sequences of 
primitive elements, this again simplifies to $O(n \lg n)$, but where the
only $O(n \lg n)$ step is the inversion count of line 26.  Therefore, for sequences
of primitive elements, such as ASCII or Unicode characters, or primitive integers
or floating-point numbers, Algorithm 2 will likely run faster than Algorithm 1.

In this analysis, we assumed that the hash table operations
are $O(1)$, which in practice should be achievable with 
sufficiently large table size and a well-designed hash function for the type of 
elements contained in the sequences.

\subsubsection{Notes on the Runtimes}

In addition to likely running faster for sequences of primitive elements, in many cases
we should expect Algorithm 2 to run faster than Algorithm 1 for sequences of elements of an object type.
Under any normal circumstances, the cost, $f_h(m)$, to compute a hash of an object of
size $m$ should be no more than linear in the size of the object.  Thus, the runtime
of Algorithm 2 should be no worse than $O(m n + n \lg n)$.  Similarly, the cost $f_c(m)$
to compare objects of size $m$ should be no worse than linear in the size of the objects.
Thus, the runtime for Algorithm 1 is no worse than $O(m n \lg n)$, which is higher order than
the runtime of Algorithm 2.  However, it is possible that a comparison of objects of size $m$
may run faster than a hash of an object of size $m$ since a comparison may short circuit
on an object attribute difference found early in the comparison.  
Therefore, Algorithm 1 may be the preferred algorithm for sequences of large objects.  
We explore this experimentally in the next section.

Furthermore, the runtime, $O(f_h(m)\, n + n \lg n)$, of Algorithm 2 is no worse than 
the runtime, $O(f_c(m)\,  n \lg n)$, of Algorithm 1 provided 
that $\frac{f_h(m)}{f_c(m)} = O(\lg n)$.  So any advantage Algorithm 1 may have on
sequences of large objects diminishes for large sequence lengths.

\section{Experiments}\label{sec:experiments}

In this section, we experimentally explore the relative performance of the two
algorithms for computing Kendall tau sequence distance.
In Section~\ref{sec:reference} we describe our reference implementations of the
two algorithms, and explain our experimental setup in Section~\ref{sec:setup}.
Then, in Section~\ref{sec:primitives}, we experimentally compare the two algorithms
on sequences of primitive values, such as strings of Unicode characters, arrays of integers,
and arrays of floating-point values. Section~\ref{sec:objects} compares the 
performance of the algorithms on arrays of objects of varying sizes..

\subsection{Reference Implementations in Java}\label{sec:reference}

We provide reference implementations of both algorithms from the previous section 
in an open source Java library available at: \url{https://jpt.cicirello.org}.
Specifically, the class KendallTauSequenceDistance, in the package
org.cicirello.sequences.distance, implements both algorithms.  
The implementations support computing the 
Kendall tau sequence distance between Java String objects, 
arrays of any of Java's primitive types
(i.e., char, byte, short, int, long, float, double, boolean), as well as
computing the distance between arrays of any object type.  

For arrays of objects,
the implementation of Algorithm 1 requires the objects to be of a class that implements
Java's Comparable interface, since the sort step requires comparing pairs of elements for
relative order; while Algorithm 2 requires the objects to be of a class
that overrides the hashCode and equals methods of Java's Object class since it relies on
a hash table.  

To compute the distance between arrays of objects,
our implementation of Algorithm 2 uses Java's HashMap class for the
hash table, and the default maximum load factor of 0.75.  
To eliminate the need to rehash to maintain that load factor, we initialize the
HashMap's size to $\lceil \frac{n}{0.75} \rceil$, where $n$ is the sequence length.
In this way, even if every element is unique, no rehashing will be needed.

For computing the distance between arrays of primitive values, as well as
for computing the distance between String objects, our implementation of Algorithm 2 
uses a set of custom hash table classes (one for each primitive type).  All of these hash tables
(except the one for bytes) use chaining with single-linked lists for the buckets.  
The size of the hash table is set, as above, based on the length of the 
array to ensure that the load factor is no higher than 0.75.  Additionally, we use
a table size that is a power of two to enable using a bitwise-and operation rather than
a mod to compute indexes.  However, we limit the table size to no greater than $2^{16}$ for the 
two 16-bit primitive types (char and short), and to no greater than $2^{30}$ for all other
types.     
The integer primitive types are hashed in the obvious way for each
of the three such types that use 16 to 32 bits (char, short, int).
Specifically, char and short values are cast to 32-bit int values.
We hash long values with an xor of the right and left 32-bit halves.  
We hash a float using its 32 bits as an int.  
We hash a double with an xor of its left and right 32-bit halves, using the
result as a 32-bit int.  Java's Float and Double classes 
provide methods for converting the bits of float and double values 
to int and long values, respectively. 
We otherwise do not use Java's wrapper classes for the primitive types.

In the case of arrays of bytes, our implementation of Algorithm 2 uses a simple
array of length 256 as the hash table, one cell for each of the possible byte values,
regardless of byte sequence length.  In this way, there are never any hash collisions when
computing the distance between arrays of byte values.

For arrays of booleans, we handle the mapping to integers differently regardless of algorithm choice,
since it is straightforward to map all false values to 0 and all true values to 1 in linear time. 

The KendallTauSequenceDistance class can be configured to use either of the two algorithms.
The default is Algorithm 2, since as we will see in Sections~\ref{sec:primitives} 
and~\ref{sec:objects}, it is always faster for sequences of primitives and nearly always
faster for arrays of objects.

\subsection{Experimental Setup}\label{sec:setup}

Our experiments are implemented in Java 1.8, and we use the Java HotSpot 64-Bit
Server VM, on a Windows 10 PC.  Our test system has 8GB RAM, with a quad-core 
AMD A10-5700 APU processor with 3.4 GHz clock speed.

\subsection{Results on Sequences of Primitives}\label{sec:primitives}

\subsubsection{Strings}\label{sec:char}

Our first set of results is on computing Kendall tau sequence distance 
between Java String objects.  Strings in Java are sequences of 16-bit char values,
that encode characters in Unicode.

In our experiments, we consider String lengths $L \in \{ 2^8, 2^9, \ldots, 2^{17} \}$,
and alphabet size $|\Sigma| \in \{ 4^0, 4^1, \ldots, 4^8 \}$.
Note that $|\Sigma| = 4^8 = 2^{16}$ is the entire Unicode character set,
and that $|\Sigma| = 2^{8}$ is the ASCII subset of Unicode.
The alphabet $\Sigma$ is just the first $|\Sigma|$ characters of the Unicode set.
For each combination of $L$ and $|\Sigma|$, we generate 100 pairs of Strings
as follows.  The first String in each pair is generated randomly, such that each
character in the String is selected uniformly at random from the alphabet $\Sigma$.
The second String is then a randomly shuffled copy of the first String.
We compute the average CPU time to calculate Kendall tau sequence distance averaged over
the 100 random pairs of Strings.

\begin{figure}[t]
\centering
\begin{tabular}{c}
\includegraphics[scale=0.84]{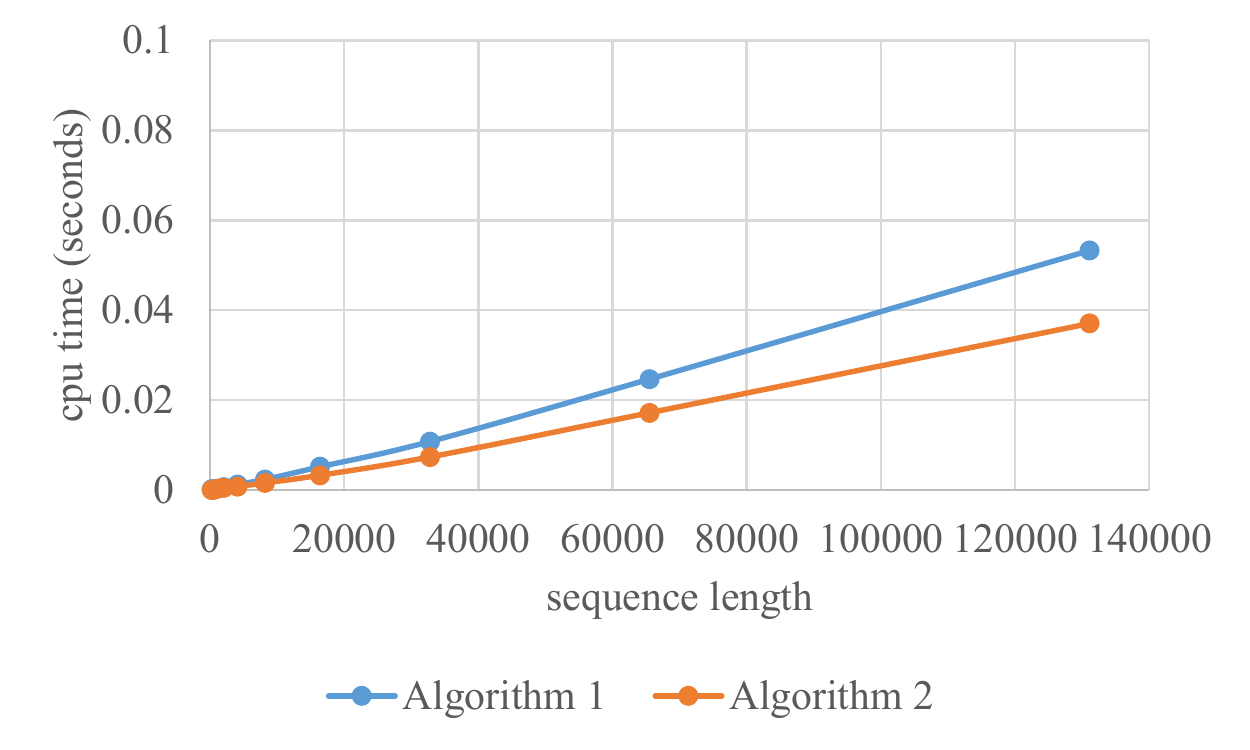} \\
(a) $|\Sigma| = 256$ \bigskip \\
\includegraphics[scale=0.84]{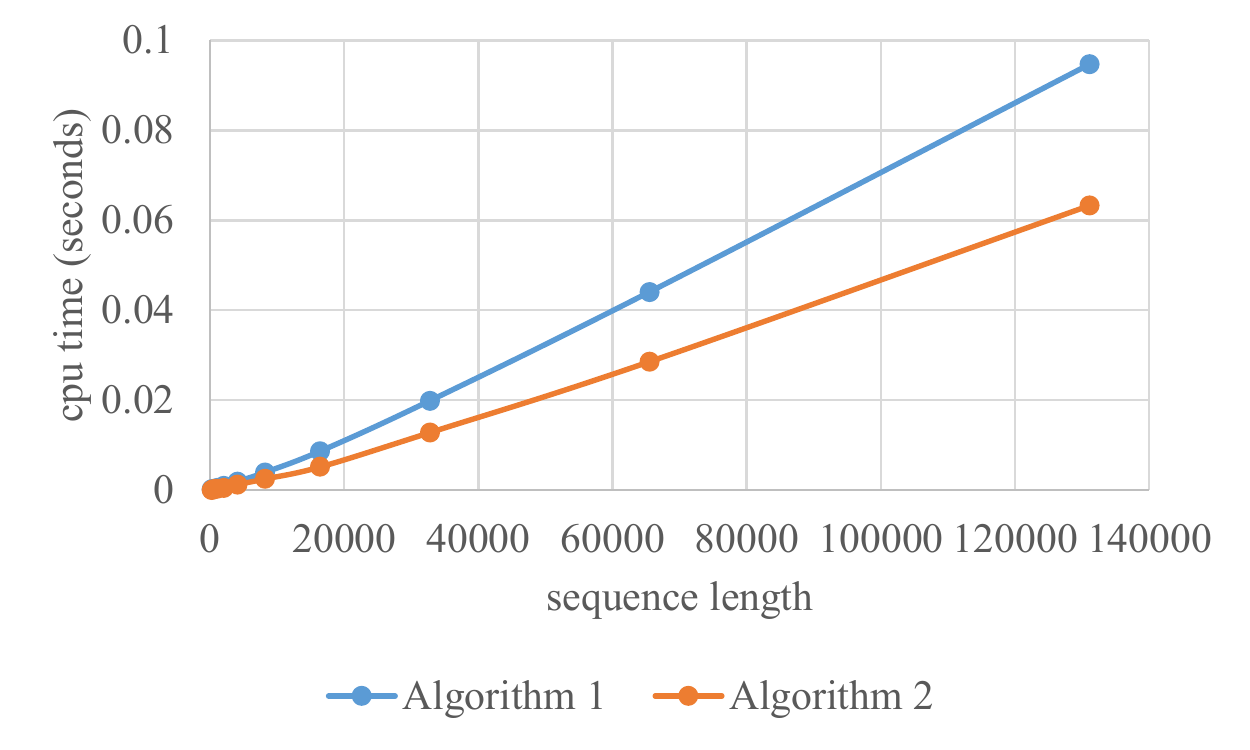} \\
(b) $|\Sigma| = 65536$
\end{tabular}
\caption{Average CPU time for Strings of characters from varying size alphabets.}\label{fig:str}
\end{figure}

Figure~\ref{fig:str} shows the results for two of the alphabet sizes: 256 and 65536.
String length is on the horizontal axis, and average CPU time is on the vertical axis.
Algorithm 2 is consistently faster than Algorithm 1, independent of alphabet size.
This is also true of the other alphabet sizes, thus we have excluded graphs in the interest of 
brevity.  The interested reader can use the code provided in the JPT repository
to replicate our experimental data.

The explanation for why alphabet size affects the runtime of the algorithms is 
straightforward.  First, note that larger alphabet size lead to longer runtime
(Figure~\ref{fig:str}(b) vs Figure~\ref{fig:str}(a)).  A smaller
alphabet size means more duplicate characters in the strings.  For Algorithm 1 that
means that the sort has fewer elements to move.  In the case of Algorithm 2, the
hash table contains one entry for each unique character in the strings, so the
smaller alphabet size leads to fewer hash table entries, which translates to
lower load factor and thus faster hash table lookups.

\subsubsection{Arrays of Integers}

\begin{figure}[t]
\centering
\begin{tabular}{c}
\includegraphics[scale=0.84]{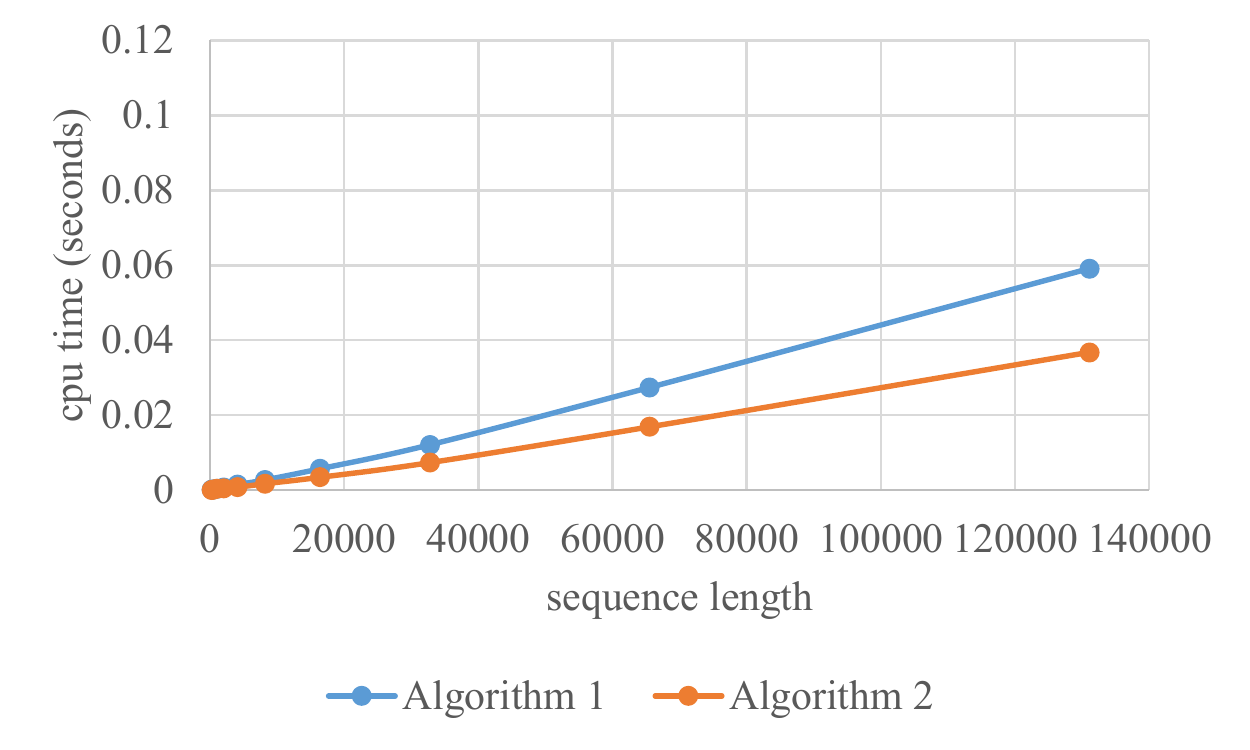} \\
(a) $|\Sigma| = 256$ \bigskip \\
\includegraphics[scale=0.84]{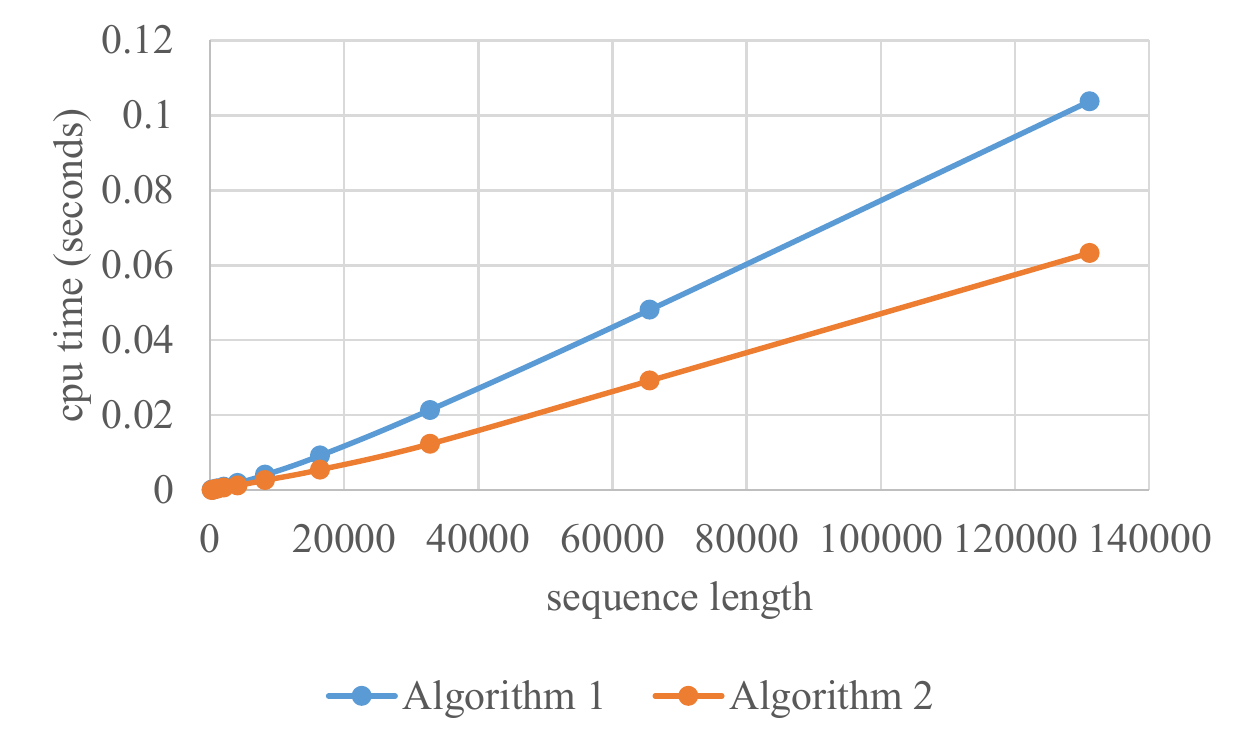} \\
(b) $|\Sigma| = 65536$
\end{tabular}
\caption{Average CPU time for sequences of 32-bit integers from varying size alphabets.}\label{fig:int}
\end{figure}

This next set of results is on computing Kendall tau sequence distance between
arrays of int values, where an int in Java is a 32-bit integer.
The array lengths $L$ are the same as the String lengths used in Section~\ref{sec:char},
as are the alphabet sizes $|\Sigma|$, where the alphabet $\Sigma$ is just the first $|\Sigma|$
non-negative integers.  We again average CPU times over 100 pairs of randomly generated
arrays, where the first array contains integers generated uniformly at random from 
the alphabet, and the second array in each pair is a randomly shuffled copy of the first.

Figure~\ref{fig:int} shows the results for two of the alphabet sizes: 256 and 65536.
Just as with Strings of characters, Algorithm 2 is consistently faster than Algorithm 1
for computing Kendall tau sequence distance between arrays of 32-bit integers, independent 
of alphabet size and array length.

Just as in the case of Strings, both algorithms run faster with the smaller alphabet size
than with a larger alphabet size.  The explanation is the same: smaller alphabet means 
more duplicate copies of elements, which means sorting is faster (Algorithm 1) and hash table 
lookups are faster due to reduced load factor (Algorithm 2).

\subsubsection{Arrays of Floating-Point Numbers}

In this last case of sequences of primitives, we consider arrays of 64-bit double-precision
floating point numbers, Java's double type.  We consider the same array lengths and alphabet
sizes as the previous cases, but now the alphabet is a set of floating-point values.  Specifically,
the alphabet $\Sigma$ contains $1.0 x$ where $x$ is the first $|\Sigma|$
non-negative integers.

\begin{figure}[t]
\centering
\begin{tabular}{c}
\includegraphics[scale=0.84]{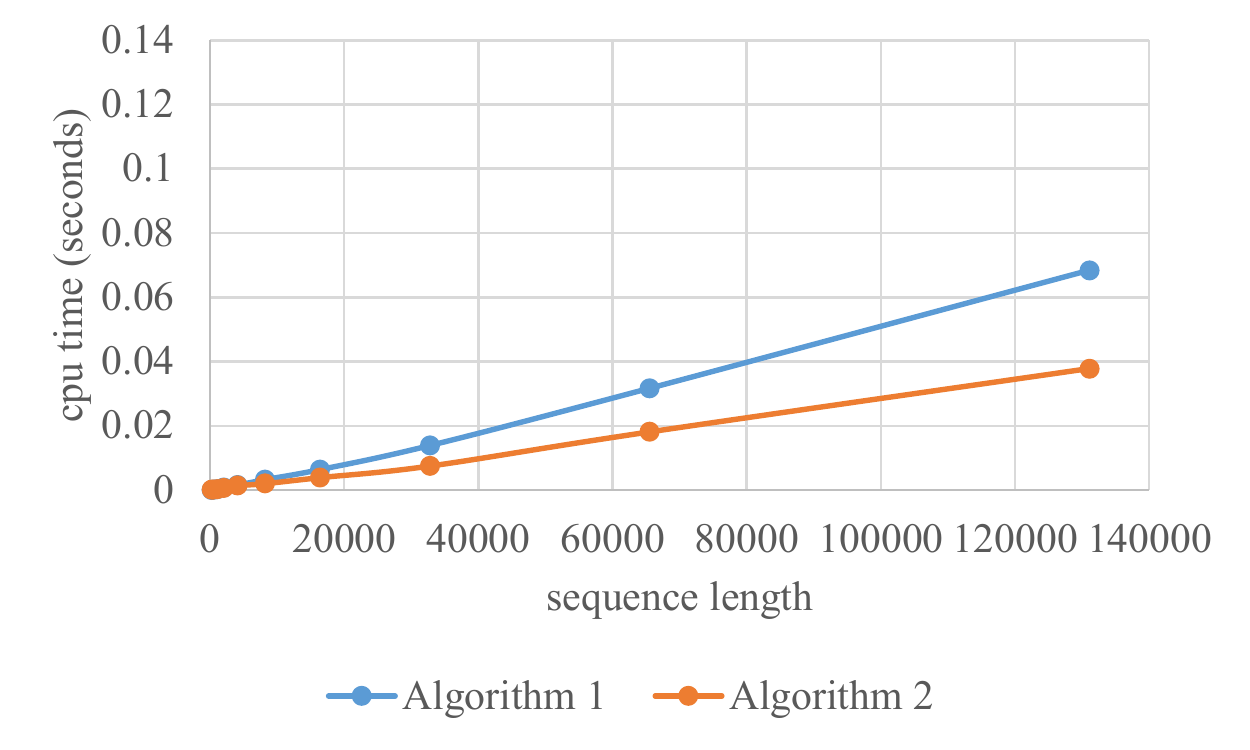} \\
(a) $|\Sigma| = 256$ \bigskip \\
\includegraphics[scale=0.84]{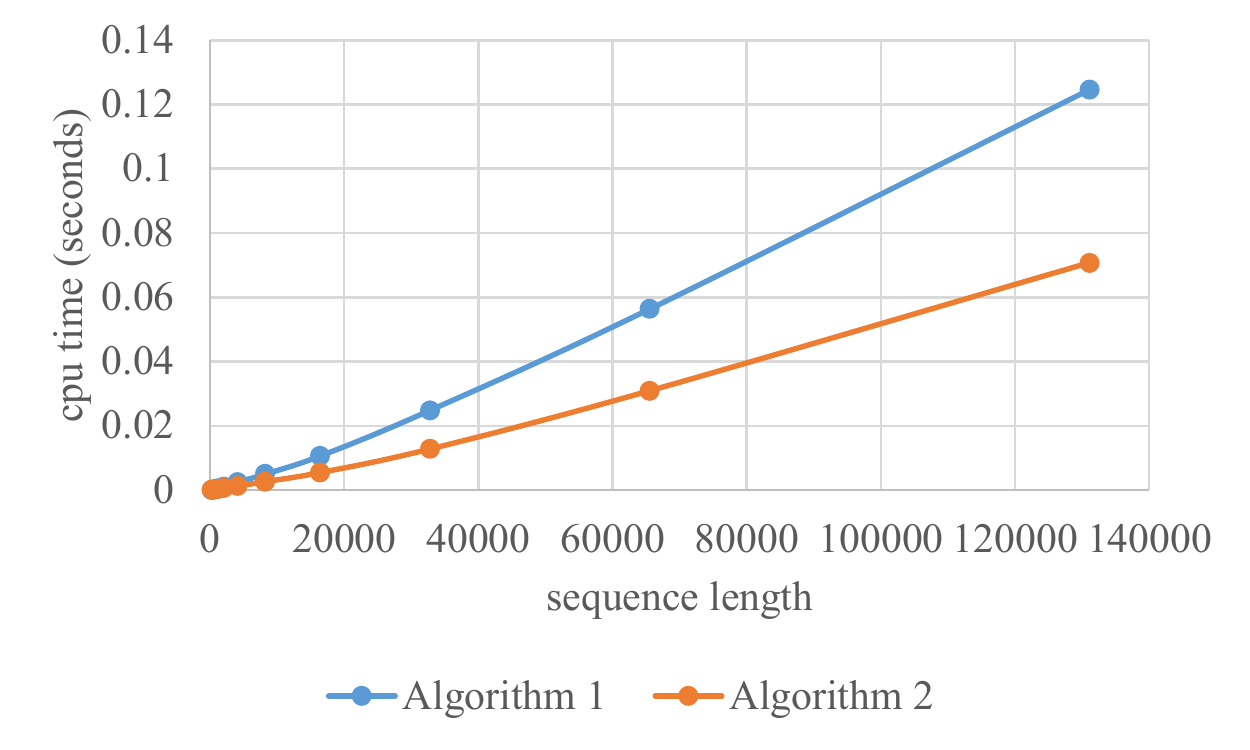} \\
(b) $|\Sigma| = 65536$
\end{tabular}
\caption{Average CPU time for sequences of 64-bit doubles from varying size alphabets.}\label{fig:double}
\end{figure}

Figure~\ref{fig:double} shows the results for two of the alphabet sizes: 256 and 65536.
Just as in the previous two cases, Algorithm 2 is consistently faster than Algorithm 1
for computing Kendall tau sequence distance between arrays of 64-bit double-precision
floating-point numbers, independent of alphabet size and array length.  And again, runtime
is longer for both algorithms with larger alphabet size for the same reasons as before.

\subsection{Results on Sequences of Objects}\label{sec:objects}

\begin{figure}[t]
\centering
\begin{tabular}{c}
\includegraphics[scale=0.84]{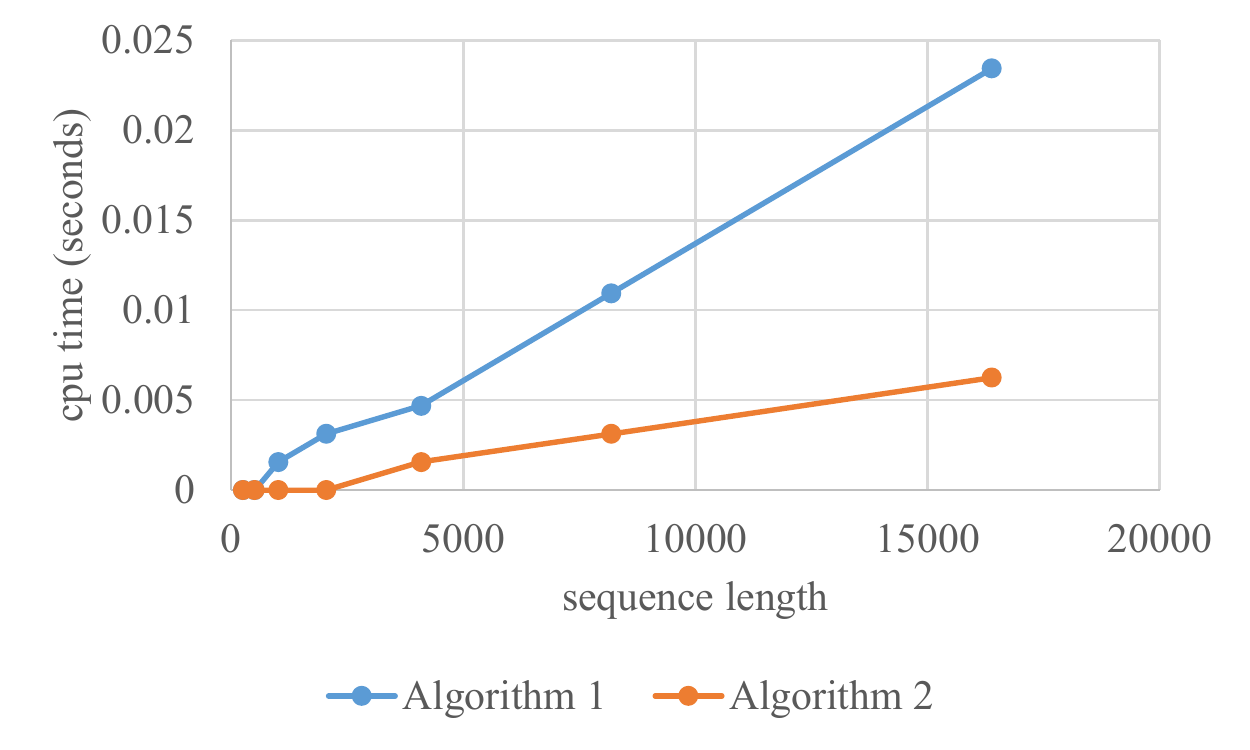} \\
(a) HCC \bigskip \\
\includegraphics[scale=0.84]{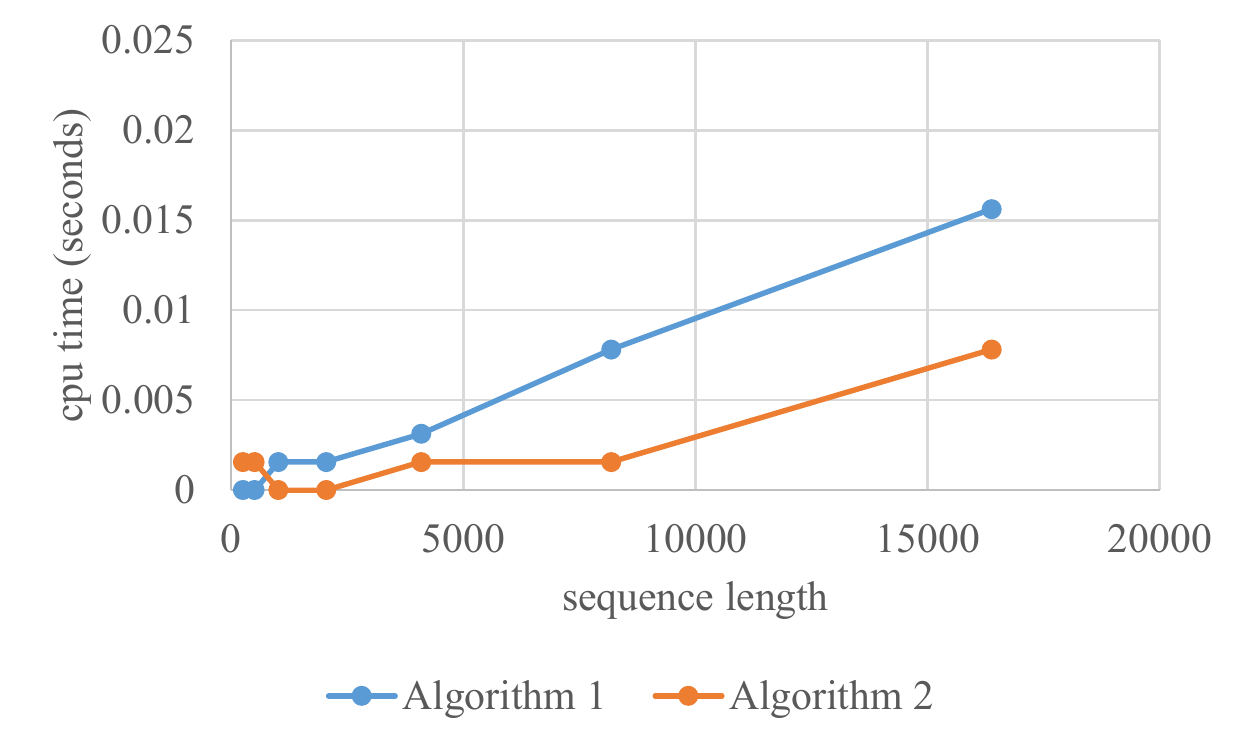} \\
(b) LCC
\end{tabular}
\caption{Average CPU time for sequences of 32 character long String objects.}\label{fig:obj32}
\end{figure}

In this section, we explore the performance of the algorithms on computing distance
between sequences of objects.  Specifically, we use arrays of Java String objects.
For example, consider sequences $s_1$ and $s_2$ as follows:
\begin{equation}
s_1 = [ ``hello'', ``world'', ``hello'', ``blue'', ``sky''] ,
\end{equation}
\begin{equation}
s_2 = [ ``hello'', ``blue'', ``sky'', ``hello'', ``world''] .
\end{equation}
These sequences are a Kendall tau sequence distance of 5 from each other.  One
sequence of adjacent swaps of length five that transforms $s_1$ into $s_2$, starts by swapping
``blue'' to the left twice, then swaps ``sky'' twice to the left, and finally swaps ``world''
with the right most of the two copies of ``hello.''

We use String objects for this set of experiments because it is easy to vary the size of a String object;
and it is also relatively easy to create a case where both a hash and a comparison have cost $O(m)$
where $m$ is object size (in this case length) as well as a case where a comparison costs significantly less
than a hash.

\begin{figure}[t]
\centering
\begin{tabular}{c}
\includegraphics[scale=0.84]{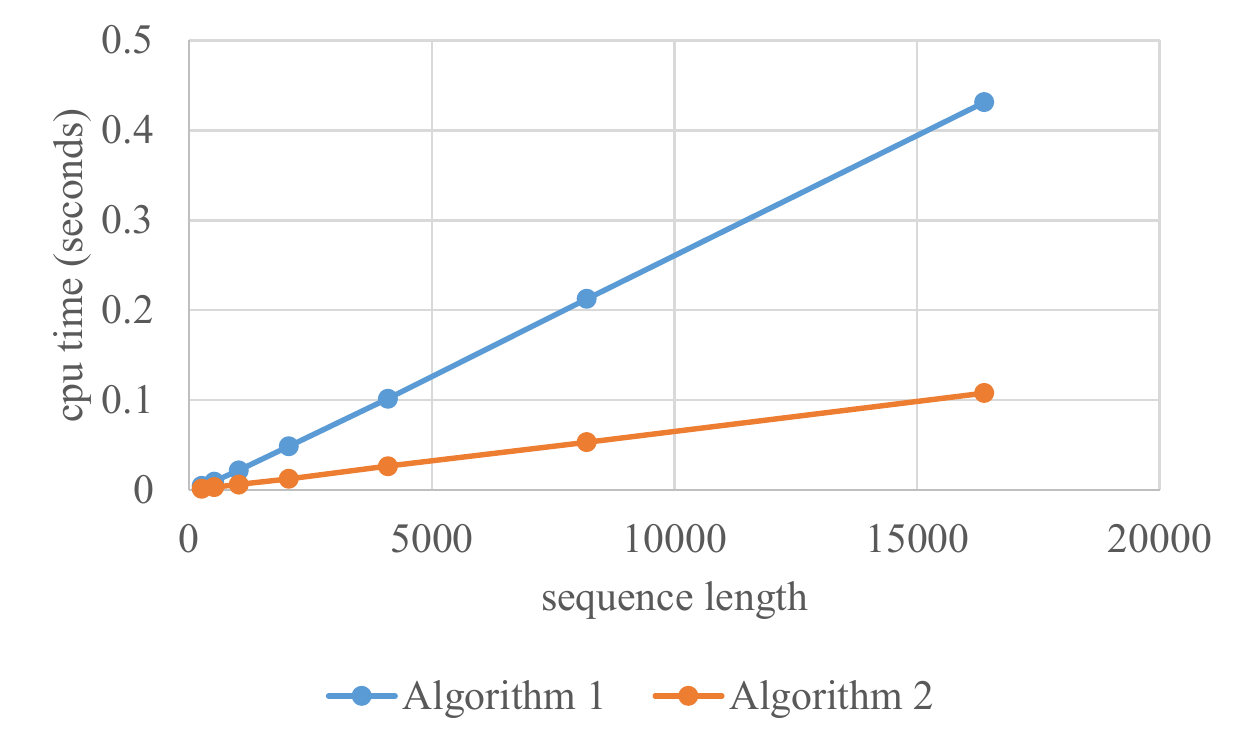} \\
(a) HCC \bigskip \\
\includegraphics[scale=0.84]{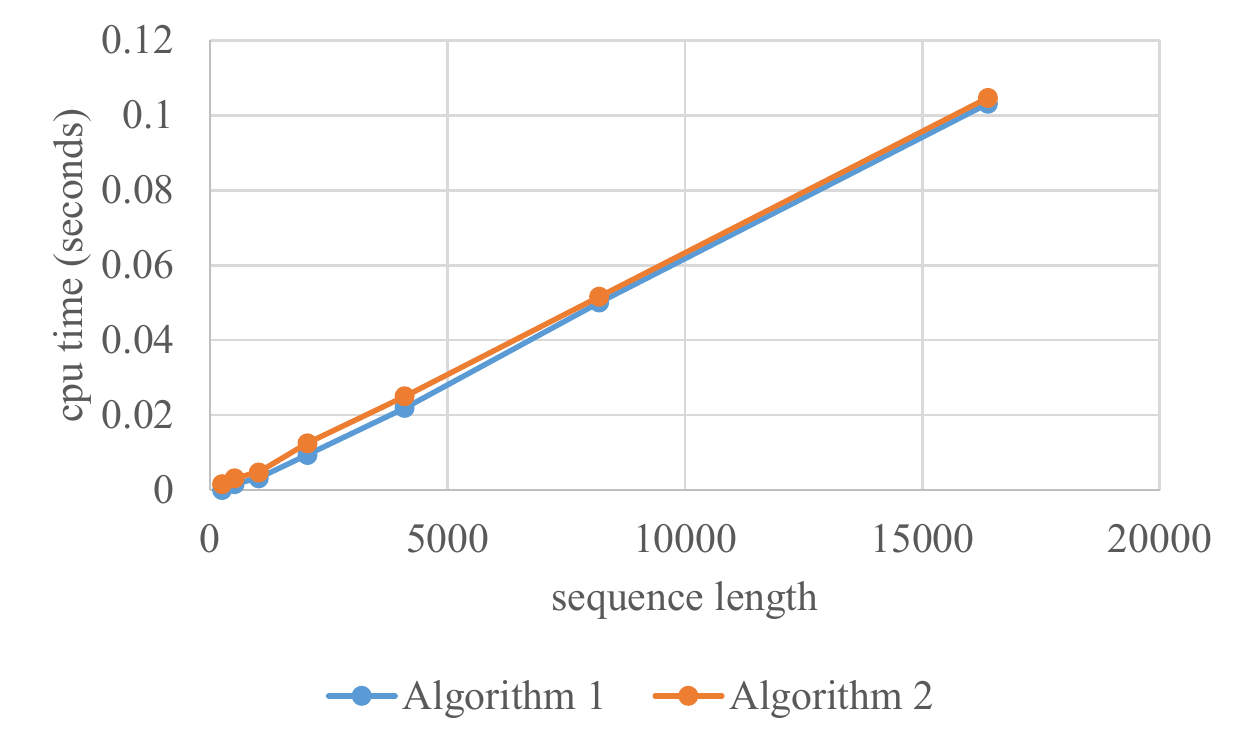} \\
(b) LCC
\end{tabular}
\caption{Average CPU time for sequences of 2048 character long String objects.}\label{fig:obj2048}
\end{figure}

We consider array lengths $L \in \{ 2^8, 2^9, \ldots, 2^{14} \}$,
and alphabet size $|\Sigma| = 256$, where the alphabet
is a set of String objects.
We consider the
following object sizes $m \in \{ 2^0, 2^1, \ldots, 2^{11} \}$.
Computing a hash of a String of length $m$ has cost $O(m)$
regardless of String content.  We consider two cases of String formation.
In the first case, each of the 256 Strings in $\Sigma$ begin with $m-1$ copies
of Unicode character 0, and only differ in the last character.  In this case, 
all comparisons also cost $O(m)$ since linear iteration over the entire String object
is required to determine how they differ.  We will refer to this case as {\em high cost
comparisons (HCC)}.
In the second case, each of the 256 Strings in $\Sigma$ is $m$ copies 
of the same character, but each of the 256 Strings use a different character.
Comparisons in this case either immediately short circuit on the first character
(if they are different) or require linear iteration if they are identical. 
We will refer to this case as {\em low cost
comparisons (LCC)}.  For each combination of $L$, $m$, and HCC vs LCC, we generate
10 pairs of sequences.  Each pair contains the same set of objects, but in different random orders.
We compute average CPU time across the 10 pairs of sequences.

In Figures~\ref{fig:obj32} and~\ref{fig:obj2048}, we show average CPU time as a function of sequence 
length for arrays of String objects 32 characters and 2048 characters in length, respectively.
Part (a) of each figure is the HCC case, and part (b) is the LCC case.
For the small objects (Figure~\ref{fig:obj32}), Algorithm 2 is consistently faster for
all sequence lengths in both the HCC and LCC cases, although the performance gap is much narrower
in the LCC case.

\begin{figure}[t]
\centering
\includegraphics[scale=0.84]{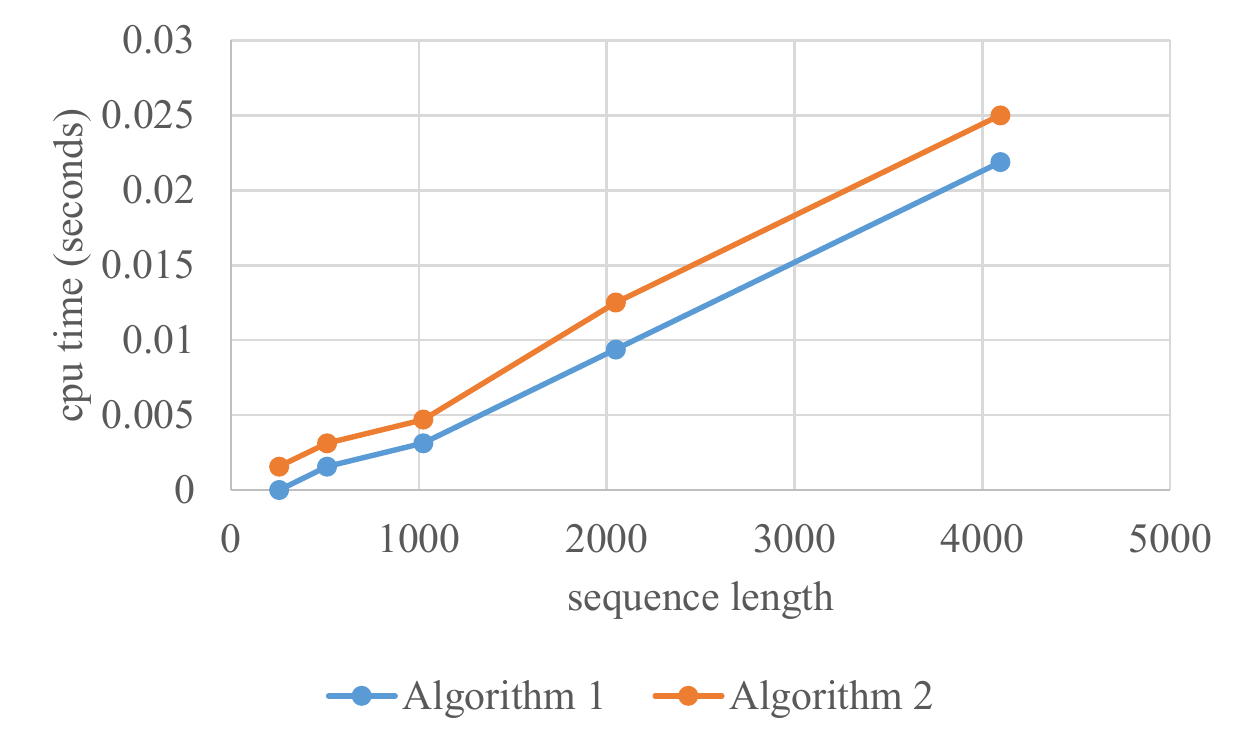} \\
\caption{Average CPU time for LCC case with sequences of 2048 
character long String objects.}\label{fig:zoom2048}
\end{figure}

For the large object case (Figure~\ref{fig:obj2048}), Algorithm 2 is faster for all
sequence lengths in the HCC case (Figure~\ref{fig:obj2048}(a)).  
For the LCC case (Figure~\ref{fig:obj2048}(b)),
when the sequence length is long, performance of the two algorithms appears to converge; 
but for shorter length sequences, Algorithm 1 is faster.  
To see this clearer, we zoom in on the left side of the graph
in Figure~\ref{fig:zoom2048}, where you can clearly see that Algorithm 1 is faster.

\section{Conclusion}

In this paper, we presented a new extension of Kendall tau distance that 
we call Kendall tau sequence distance.  The original Kendall tau distance
is a distance metric on permutations.  We have adapted it to be applicable
for computing distance between general sequences.  Both sequences must
be of the same length and contain the same set of elements, otherwise the
Kendall tau sequence distance is undefined.

We introduced two algorithms for computing Kendall tau sequence distance.
If the sequences contain primitive values, such as a string of characters,
or an array of primitive integers, etc, then the runtime of both algorithms
is $O(n \lg n)$.  However, the only $O(n \lg n)$ step of Algorithm 2 is
a permutation inversion count that is shared with Algorithm 1; and thus, 
Algorithm 2 should be preferred for sequences of primitives.
If one is computing the distance between sequences of objects
of some more complex type, then the size of the objects in the sequences also
impacts the runtime of the algorithms.  However, unless the cost of a
hash of an object is significantly greater than the cost of an object comparison,
Algorithm 2 is still the preferred algorithm.

We provide reference implementations of both algorithms in the Java language.
These implementations have been made available in an open source library.
Our experiments confirm that Algorithm 2 is the faster algorithm under 
most circumstances.  The code to replicate our experimental data is also 
available as open source.

\bibliographystyle{plainnat}
\bibliography{tau}

\end{document}